\documentclass[epj]{svjour}
\usepackage{amsmath}
\usepackage{graphicx}
\usepackage{color}
\usepackage[english]{babel}
\usepackage{mathpazo}
\newcommand{\pbard}{{\bar p} \, d \to e^+ e^- n}
\newcommand{\pbarp}{{\bar p} \, p \to e^+ e^-}
\newcommand{\mee}{{\cal M}}
\newcommand{\panda}{\overline{\mathrm{P}} \mathrm{ANDA}}
\begin{document}
%
%
\title{
{Antiproton-nucleus electromagnetic annihilation as a way to
access the proton timelike form factors}}
\author{H.~Fonvieille\inst{1} \and V.A.~Karmanov\inst{2}}
\institute{
Laboratoire de Physique Corpusculaire, Universit\'e Blaise Pascal, IN2P3, 63177 Aubi\`ere Cedex,
France \and{Lebedev Physical Institute, Leninsky Prospekt 53,
119991 Moscow, Russia} }

\date{Received: \ldots / Revised version: \ldots}

\abstract{Contrary to the reaction $\pbarp$ with a high momentum
incident antiproton  on a free target proton at rest, in which
the invariant mass ${\cal M}$ of the $e^+e^-$ pair is
necessarily much larger than the $\bar{p}p$ mass $2m$, in the
reaction $\pbard$ the value of ${\cal M}$ can take values near
or below the $\bar{p}p$ mass. In the antiproton-deuteron
electromagnetic annihilation, this allows to access the proton
electromagnetic form factors in the time-like region of $q^2$
near the $\bar{p}p$ threshold. We estimate the cross section
$d\sigma_{\bar{p}d\to e^+e^-n}/d\mee$ for an antiproton beam
momentum of 1.5 GeV/c. We find that near the $\bar{p}p$
threshold this cross section is about 1\;pb/MeV. The case of
heavy nuclei target is also discussed. Elements of experimental
feasibility are presented for the process $\pbard$ in the
context of the $\panda$ project.}
\PACS{
    {25.43.+t}{Antiproton-induced reactions}   \and
    {14.20.Dh}{Properties of protons and neutrons}             \and
    {13.40.Gp}{Electromagnetic form factors}
           }

\titlerunning{Antiproton-nucleus electromagnetic
annihilation...}

\maketitle

\section{Introduction}
\label{intro}

A lot of efforts are devoted to understanding
the structure of the nucleon, the building block of matter. The
underlying theory is the field theory of strong interaction,
QCD, in its non-perturbative regime. Electromagnetic
pro\-p\-erties of the nucleon are fundamental pieces to this
puzzle. Among them, the electromagnetic form factors of the
proton and the neutron are basic observables, which are the goal
of  extensive measurements. In the spacelike region, i.e. for a
virtual photon four-momentum squared $q^2<0$, these form factors
give information about the spatial distribution of electric
charge and magnetization inside the nucleon. In the timelike
region ($q^2>0$) they tell us about the dynamics of the
nucleon-antinucleon ($N \bar N$) interaction.

\begin{figure}[h!]
\begin{center}
\includegraphics[width=8cm]{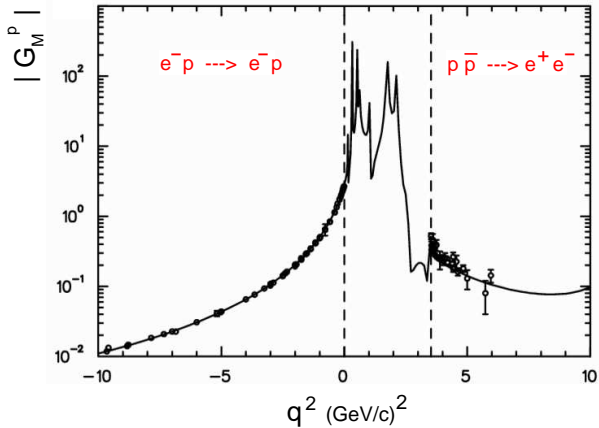}
\caption{Experimental data and predictions for the magnetic
proton form factor in the domain $-10\;GeV^2/c^2\le q^2 \le
10\;GeV^2/c^2$. The figure is taken from
\cite{Meissner:1999hk}.
\label{fig1}}
\end{center}
\end{figure}

A fully consistent description of the nucleon form factors
should include both domains, of spacelike and timelike $q^2$,
since these domains are related by crossing symmetry. Such
theoretical models  are generally based on  dispersion relations
\cite{Baldini:1998qn,Hammer:1996kx,Adamuscin:2005aq} or
semi-phenomenological approaches
\cite{Iachello:2004aq,Bijker:2004yu}. They predict a smooth
behavior of the form factor in the measured regions,  but a
peaked behavior in the timelike region below the $N \bar N$
threshold ($0 < q^2 < 4m^2$, where $m$ is the nucleon mass), due
to poles in the amplitude (see e.g. fig.~\ref{fig1}, taken from
\cite{Meissner:1999hk}). These poles are phenomenological
inputs, built from meson exchange,  and their properties are
fitted to the data in the measured regions. The corresponding
irregularities in form factors are related to the transition of
$p\bar{p}$ to vector mesons which can decay in $e^+e^-$ pair via
a virtual photon.

The mesons with a  mass near the $p\bar{p}$ mass can have a
quasinuclear nature, i.e., they can be formed by bound states
and resonances in the $p\bar{p}$ system. Such vector mesons were
predicted in the papers~\cite{Dalkarov:1989xs,Dalkarov:1989dc}.
Note that such mesons can be formed not only in the $p\bar{p}$
system but in $N\bar{N}$ in general and they can  have not only
vector quantum numbers. A review on quasinuclear mesons in the
$N\bar{N}$ system is given in \cite{Shapiro:1978wi}.

The under-threshold region ($0 < q^2 < 4m^2$)
is called unphysical because it cannot be accessed
experimentally by an on-shell process. Some experiments have
been performed in the vicinity of the $N \bar N$ threshold,
either in $p \bar p \to e^+ e^-$  at LEAR \cite{Bardin:1994am}
or in the inverse channel $e^+ e^- \to p \bar p$ at Babar
\cite{Aubert:2005cb}, but they cannot go below this physical
threshold. However, a nucleus provides nucleons
with various momenta, in modulus and direction, and also various
degrees of off-shellness.
Therefore it offers the possibility to produce
an $N \bar N$ electromagnetic annihilation with an invariant
mass squared $q^2= s_{\bar p p}$ smaller than $4 m^2$.  The main
purpose of our paper is to explore this possibility, which may
give access to the proton form factors in the underthreshold
region, for an off-shell nucleon. The idea to use a nucleus for
that purpose was explored in the 80's using deuterium
~\cite{Dalkarov:1980de}. The reaction is then:
\begin{equation}\label{(1)}
\pbard
\end{equation}
(a crossed-channel of deuteron electrodisintegration). The aim of
the present paper is to revive this study in view of the future
antiproton facility FAIR at GSI.

Other channels can give access to the off-shell nucleon form
factors in the timelike region, including the underthreshold
region; such processes have been studied theoretically in
ref.~\cite{Schafer:1994vr} ($\gamma p \to p e^+ e^-$) and in
refs.~\cite{Adamuscin:2007iv,Dubnickova:1995ns}
 (${\bar p} p \to \pi^0 e^+ e^-$).

The paper is organized as follows: a theoretical study is
presented in sect.~\ref{sec-globalth} and experimental aspects
are presented  in sect.~\ref{sec-globalexp}. 
Other aspects are mentioned in sect.~\ref{sec-other} and a conclusion
is given in sect.~\ref{sec-concl}.


\section{Theoretical study}
\label{sec-globalth}

In elastic electron scattering from the nucleon  $e^-N\to e^-N$
the momentum transfer squared $q^2=(k-k')^2$ is always negative.
This allows to measure the nucleon form factors in the
space-like domain of $q^2$.

On the contrary, in the annihilation $N\bar{N}\to \gamma^*\to
e^+e^-$ the mass of virtual photon is equal to the total c.m.
$N\bar{N}$ energy. Its four-momentum squared is always greater
than $4m^2$. This allows to measure the nucleon form factors in
the time-like domain of $q^2$, above the $N\bar{N}$ threshold.
In this reaction, in order to study the form factor behavior in
a narrow domain near threshold, where non-trivial structures are
predicted \cite{Shapiro:1978wi}, one should have a beam of
almost stopped antiprotons. This non-easy technical problem was
solved at LEAR ~\cite{Bardin:1994am}. However, the
under-threshold  domain $0\le q^2 \le 4m^2$ remains
kinematically unreachable in this type of experiments.

\begin{figure}[h!]
\begin{center}
\includegraphics[width=8cm]{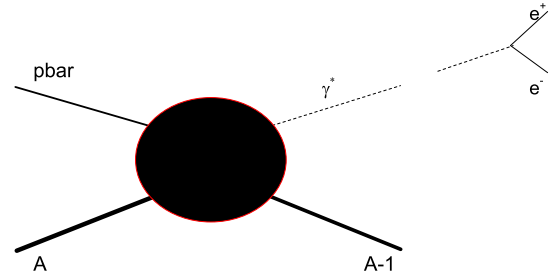}
\end{center}
\caption{The process $\bar{p}A\to (A-1)\gamma^*$  (followed
by $\gamma^*\to e^+e^-)$.}
\label{pbarA}
\end{figure}

\begin{figure}[h!]
\begin{center}
\includegraphics[width=8cm]{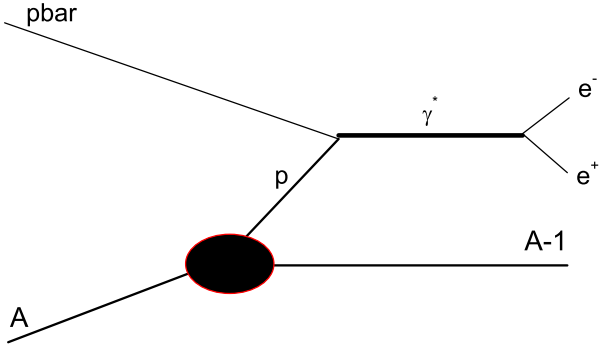}
\end{center}
\caption{Amplitude of the reaction $\bar{p}A\to (A-1)\gamma^*$
in impulse approximation. }
\label{eAMX}
\end{figure}

One can penetrate in this domain of $q^2$ in the $\bar{p}$
annihilation on nuclei
$$\bar{p}A\to (A-1) \ e^+e^-,$$
see fig. \ref{pbarA}. The symbol $(A-1)$ means not necessarily a
nucleus but any system with the baryon number $A-1$. Since extra
energy of the antiproton can be absorbed by the $(A-1)$ system,
the $e^+e^-$ pair may be emitted with very small invariant mass.
Therefore the two-body reaction $\bar{p}A\to (A-1)\gamma^*$ is
kinematically allowed for a very wide domain of invariant mass
of the $\gamma^*$, which starts with two times the electron
mass, namely:
$$4m_e^2\le q^2 \le (\sqrt{s_{\bar{p}A}}-M_{A-1})^2 \ . $$
One can achieve near-threshold, under-threshold and even
deep-under-threshold values of $q^2$ even for fast antiprotons.
This however does not mean that this reaction provides us direct
information about the nucleon form factors. For the latter, we
should be sure that the observed $e^+e^-$ pair (and nothing
more) was created in the annihilation $\bar{p}p\to e^+e^-$ on
the proton in the nucleus, i.e., that the reaction mechanism is
given by the diagram of fig. \ref{eAMX} or by a similar diagram
where the $\bar{p}$ can rescatter before annihilation.

\begin{figure}[ht!]
\begin{center}
\includegraphics[width=6cm]{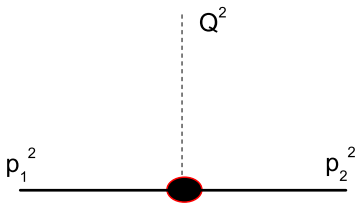}
\end{center}
 \caption{Nucleon electromagnetic vertex.}
\label{NFF}
\end{figure}

At the same time, since the nucleons in the nucleus are
off-mass-shell, the form factors  entering the
amplitude of fig. \ref{eAMX}, are not precisely the same as
found in the free $\bar{p}p$ annihilation. In general, the
three-leg vertex, shown in fig. \ref{NFF}, depends not only on
the photon virtuality $q^2$, but also on the nucleon ones
 $p^2_1,p^2_2$: $F=F(p^2_1,p^2_2,q^2)$. In the
case considered, the incident antiproton is on-ener\-gy-shell:
 $p^2_{\bar{p}}=m^2$, however the form
factors depend on the proton virtuality $p^2_p$. How the form
factor $F(p^2_{\bar{p}}=m^2,p^2_p\neq m^2,q^2)$ v.s. $q^2$
differs from the free one $F(p^2_{\bar{p}}=m^2,p^2_p= m^2,q^2)$
-- this depends on the dynamics determining its behavior v.s.
the nucleon leg virtuality. The nucleon form factors with
off-shell nucleons were studied in the papers
\cite{Naus:1987kv,Tiemeijer:1990zp}.
When only one nucleon is off-shell, then there are six form
factors instead of two in the on-shell case. Generally, we can
expect that the form factor dependence v.s. $p^2_p$ is much
smoother than the $q^2$ dependence. The $p^2_p$ dependence can
be determined by the nucleon self-energy corrections (i.e., by
the structure of the nucleon), whereas the $q^2$ dependence in
the time-like domain is governed  by the $\bar{p}p$ interaction.
The  nucleon dynamics has a much larger energy scale than the
nuclear one. The typical off-shell variation found in the papers
\cite{Naus:1987kv,Tiemeijer:1990zp}
was from a few to 10 percent. We do not pretend to such an
accuracy here. Therefore we neglect this effect in our
calculation. We will come to this question later. In any case,
both domains: $q^2<4m^2,\; p^2_p=m^2$ and $q^2<4m^2,\;
p^2_p<m^2$ are totally unexplored experimentally and are
interesting and intriguing.

We emphasize that though the form factor dependence on $p^2_p$
can be weak, the nucleon off-mass-shell effect is very important for
the kinematical possibility to reach the near- and
under-threshold domain of $q^2$ with fast antiprotons. To
produce the near-threshold $e^+e^-$ pairs in annihilation of a
fast $\bar{p}$ on an on-mass-shell proton, the antiproton should
meet in the nucleus a fast proton with parallel momentum. The
probability of that, which we estimated in the Glauber approach,
is negligibly small~\cite{Dalkarov:2008}
relative to the results presented below. However, if the
effective mass $p^2_p$ of the virtual proton is smaller than
$m^2$ (that is just the case in a nucleus), then the near- and
under-threshold $e^+e^-$ pairs can be produced in collisions
with not so fast intra-nucleus nucleons. This effect
considerably increases the cross section. To have an idea of the
order of magnitude which one can expect for this cross section,
we will calculate it in the impulse approximation. Numerical
applications will be done for the lowest antiproton beam
momentum foreseen in future projects. Namely,  at the High
Energy Storage Ring at FAIR-GSI this value is 1.5 GeV/c.


\subsection{Cross section calculation}
\label{sec:theocross}

At first, we consider the case of the
deuteron target. If we know the amplitude of the reaction
$\bar{p}d\to e^+e^-n$: $M_{\bar{p}d \to e^+e^-n}$ (to be
calculated below), then the corresponding cross section is given
by:
\begin{eqnarray}\label{dsig1}
&&d\sigma_{\bar{p}d\to e^+e^-n} =\frac{(2\pi)^4}{4I}|M_{\bar{p}d
\to e^+e^-n}|^2
\nonumber\\
&&\times \delta^{(4)}(p_{\bar{p}}+p_d-p_{e^+}-p_{e^-}-p_n)
\nonumber\\
&&\times\frac{d^3 p_{e^+}}{(2\pi)^3 2\epsilon_{e^+}}\;\frac{d^3
p_{e^-}}{(2\pi)^3 2\epsilon_{e^-}}\; \frac{d^3 p_n}{(2\pi)^3
2\epsilon_n}
\end{eqnarray}
where $I$ results from the flux factors:\\
$I=j\epsilon_{\bar{p}} \epsilon_{d} =[(p_{\bar{p}}\cdot
p_d)^2-m^2M_d^2]^{1/2}=M_d p_{\bar{p},lab}\approx 2m
p_{\bar{p},lab}$, $p_{\bar{p},lab}$ is the incident $\bar{p}$
momentum in the lab. system. Here and below we imply the sum
over the final spin projections and average over the initial
ones.

We are interested in the distribution  in the invariant mass
${\cal M}$ of the final $e^+e^-$ system. To find it, for fixed
value of ${\cal M}$, we can integrate, in some limits, over the
angles of the recoil neutron (determining the neutron recoil
momentum) and over the angles of the emitted $e^+e^-$ in
their center of mass. This can be done using standard techniques
of the phase volume transformations. Namely, we  use the
identity:
$$
\int \delta^{(4)}({\cal P}_{e^+e^-}-p_{e^+}-p_{e^-})d^4 {\cal
P}_{e^+e^-} \delta({\cal P}^2_{e^+e^-}-\mee^2)d\mee^2\equiv 1,
$$
where ${\cal P}_{e^+e^-}$ is the four-momentum of the $e^+e^-$
pair and $\mee^2$ is its invariant mass squared. We insert this
formula in the cross section (\ref{dsig1}) and represent the
three-body phase volume in (\ref{dsig1}) as:
\begin{eqnarray*}
&&d\sigma_{\bar{p}d\to
e^+e^-n}=\int\ldots\delta^{(4)}(p_{\bar{p}}+p_d-p_{e^+}-p_{e^-}-p_n)
\\
&&\times\frac{d^3 p_{e^+}}{ 2\epsilon_{e^+}}\;\frac{d^3
p_{e^-}}{ 2\epsilon_{e^-}}\;  \frac{d^3 p_n}{ 2\epsilon_n}
\\
&&=\int\ldots\delta^{(4)}(p_{\bar{p}}+p_d-{\cal P}_{e^+e^-}-p_n)
\\
&&\times\delta^{(4)}({\cal P}_{e^+e^-}-p_{e^+}-p_{e^-})d^4 {\cal
P}_{e^+e^-} \delta({\cal P}^2_{e^+e^-}-\mee^2)d\mee^2
\\
&\times& \frac{d^3 p_n}{2\epsilon_n}\; \frac{d^3 p_{e^+}}{
2\epsilon_{e^+}} \;\frac{d^3 p_{e^-}}{ 2\epsilon_{e^-}}
=\int\ldots dV_1\, dV_2\, d\mee^2
\end{eqnarray*}
where $\epsilon_n$ is the energy corresponding to the
four-mo\-m\-e\-n\-tum $p_n$ and similarly
for other energies, and we give explicitly
only the factors resulting from the phase volume. We denoted the
phase volume of the $e^+e^-$ pair as:
\begin{eqnarray}
dV_1&=&\int \delta^{(4)}({\cal P}_{e^+e^-}-p_{e^+}-p_{e^-})\;
\frac{d^3 p_{e^-}}{2\epsilon_{e^-}}\; \frac{d^3 p_{e^+}}{
2\epsilon_{e^+}}
\nonumber\\
&=&\frac{p^*_{e}}{4\mee}\;d \Omega_{e}\approx\frac{1}{8}
\;d\Omega_{e} \ .
\label{V1}
\end{eqnarray}
Here $p^*_{e}$ is the final electron (or positron) momentum in
the c.m. frame of the $e^+e^-$ pair and $\Omega_{e}$ is its
solid angle in this frame. Neglecting the electron mass, we
replaced: $p^*_{e}\approx \frac{1}{2}\mee$.

The two-body phase volume $dV_2$ of the final $n\gamma^*$ state
reads:
\begin{eqnarray}
dV_2&=&\int\delta^{(4)}(p_{\bar{p}}+p_d-{\cal P}_{e^+e^-}-p_n)\;
\frac{d^3 {\cal P}_{e^+e^-}}{2\epsilon_\mee} \;\frac{d^3
p_n}{2\epsilon_n}
\nonumber\\
&=&\frac{p^*_{\gamma^* n}}{4\sqrt{s_{\bar{p}d}}}\;d
\Omega^*_{\gamma^* n} \ .
\label{V2}
\end{eqnarray}
Here $p^*_{\gamma^* n}$ and $\Omega^*_{\gamma^* n}$ are in the
c.m. frame of the reaction (i.e., the c.m. frame of $\bar{p}d$)
and $s_{\bar{p}d}$  is the corresponding c.m. energy squared.

In this way we obtain
\begin{equation}\label{sigma1}
d\sigma_{\bar{p}d\to e^+e^-n}=\frac{1}{2^8\pi^5m p_{\bar p, lab
}}\int|M_{\bar{p}d \to e^+e^-n}|^2 dV_1 dV_2\;d \mee^2 \ .
\end{equation}
This form of the cross section will provide us the invariant
mass distribution $d\sigma_{\bar{p}d\to e^+e^-n}/d \mee$.

The formula (\ref{sigma1}) is general and does not assume any
particular mechanism of the reaction. In the impulse
approximation, when the mechanism is given by the diagram of
fig. \ref{eAMX}, the total amplitude squared $|M_{\bar{p}d \to
e^+e^-n}|^2$ is proportional to the annihilation amplitude
squared\\ $|M_{\bar{p}p \to e^+e^-}|^2$ and to the square of the
deuteron wave function $|\psi|^2$. The direct calculation of the
amplitude corresponding to the diagram of fig. \ref{eAMX} gives
the proportionality coefficient:
\begin{equation}\label{Mbar}
|M_{\bar{p}d \to e^+e^-n}|^2 = 4m\,|M_{\bar{p}p \to
e^+e^-}|^2\;|\psi|^2, \
\end{equation}
and $|\psi|^2$ is normalized as:
\begin{equation}\label{norm0}
\int|\psi(k)|^2\frac{md^3k}{(2\pi)^3\epsilon_k}=1.
\end{equation}
Substituting (\ref{Mbar}) into (\ref{sigma1}), we find the cross
section:
\begin{eqnarray*}
d\sigma_{\bar{p}d \to e^+e^-n}&=&\frac{1}{2^6\pi^5
p_{\bar{p},lab}}\int |M_{\bar{p}p \to e^+e^-})|^2 \frac{1}{8}
\;d\Omega_{e}
\\
&&\phantom{\frac{1}{
p_{\bar{p},lab}}\int} \times |\psi|^2\frac{p^*_{\gamma^*
n}}{4\sqrt{s_{\bar{p}d}}}\;d \Omega^*_{\gamma^* n}\; d\mee^2 \ .
\end{eqnarray*}
We integrate over $d\Omega_{e}$ (in the $e^+e^-$ c.m. system) and
finally obtain:
\begin{equation}\label{sig2}
\frac{d\sigma_{\bar{p}d \to e^+e^-n}}{d\mee}=\sigma_{\bar{p}p
\to e^+e^-}(\mee)\;\eta(\mee),
\end{equation}
where $\eta(\mee)$ is  the distribution (given by eq.
(\ref{sig3}) below) of the $e^+e^-$ invariant mass $\mee$ due to
the fact that the reaction occurs on the proton bound in the
nucleus, and $\sigma_{\bar{p}p \to e^+e^-}(\mee)$ is the cross
section of the $\bar{p}p \to e^+e^-$ annihilation at the total
energy $\mee$. The latter is determined by the amplitude
$M_{\bar{p}p \to e^+e^-}$, corresponding to annihilation via the
$s$-channel photon $\gamma^*$. The calculation of this amplitude
is standard. Though the target proton is off-mass-shell, we do
not consider the off-mass-shell effects in this amplitude, i.e. we
assume a free proton. Then the amplitude squared reads:
\begin{eqnarray}\label{Mbar2}
|M_{\bar{p}p \to e^+e^-}|^2&=& \frac{16\alpha^2\pi^2}{\mee^2}
\cdot [  4 m^2  \vert G_E \vert ^2 \sin^2 \theta_{e^+}
\nonumber\\
&+&  \mee^2  \vert G_M \vert ^2 \cos^2\theta_{e^+} ] ,
\end{eqnarray}
where $\theta_{e^+}$ is the emission angle of the $e^+$ or the
$e^-$ in the $\gamma^*$ c.m. frame and $G_E, G_M$ are the proton
electric and magnetic timelike form factors. To estimate the nuclear
effect, from now on we omit the form factors, i.e. we put $\vert
G_E \vert = \vert G_M \vert = 1$, or $\vert F_1 \vert =1, \vert
F_2 \vert =0$ as for a pointlike nucleon. Then the $\bar{p}p \to
e^+e^-$ cross section obtains the form:
\begin{eqnarray}\label{sig4}
\sigma_{\bar{p}p \to e^+e^-}&=& \frac{1}{64 \pi^2}\frac{p^*_e}{m
p_{\bar{p},lab}\mee}\int |M_{\bar{p}p \to e^+e^-}|^2 d\Omega_{e}
\nonumber\\
&=&\frac{2\alpha^2\pi(2m^2+\mee^2)}{3\mee^2m p_{\bar{p},lab}} \ .
\end{eqnarray}

Though $\sigma_{\bar{p}p \to e^+e^-}$ depends on
$\mee$, the main (nuclear) effect is determined by the factor
$\eta(\mee)$:
\begin{eqnarray}\label{sig3}
\eta(\mee)&=&\frac{m p^*_{\gamma^*
n}\mee}{(2\pi)^2\sqrt{s_{\bar{p}d}}}\int_{-1}^1|\psi(k)|^2dz.
\end{eqnarray}
Here $z=\cos\theta$, where $\theta$ is the angle, in the c.m.
frame of the reaction, between the initial deuteron momentum
$\vec{p}^*_d$ and the final neutron momentum $\vec{p}^*_n$.
The integration over $dz$ in $\eta(\mee)$
results from $d \Omega^*_{\gamma^* n}$ in the phase volume
$dV_2$, eq. (\ref{V2}). The argument of the wave function $k$
depends on $z$. This explicit dependence is given in the next
section.

The distribution $\eta(\mee)$ is normalized to 1:
\begin{equation}\label{norm}
\int_0^{\infty}\eta(\mee)d \mee= 1.
\end{equation}
To prove eq. (\ref{norm}), we calculate this integral
explicitly:
$$
\int_0^{\infty}\eta(\mee)d \mee=\int|\psi(k)|^2\frac{m
p^*_{\gamma^* n}\mee}{(2\pi)^2\sqrt{s_{\bar{p}d}}}d
\mee\sin\theta d\theta \frac{d\phi}{2\pi} \, .
$$
Since
\begin{equation}\label{En}
\epsilon^*_n=(s_{\bar{p}d}-\mee^2+m^2)/(2\sqrt{s_{\bar{p}d}}),
\end{equation}
we have: $\mee\,d \mee=-\sqrt{s_{\bar{p}d}}\,d\epsilon^*_n$.
Then: $d\epsilon^*_n=\ \frac{p^*_ndp^*_n}{\epsilon^*_n}$. In
this way we find:
\begin{eqnarray*}
\int\eta(\mee)d \mee&=&\int|\psi(k)|^2\frac{m
p^{*2}_ndp^*_n}{(2\pi)^3\epsilon^*_n}d\Omega^*
\\
&=& \int|\psi(k)|^2\frac{md^3k}{(2\pi)^3\epsilon_k}=1 \ .
\end{eqnarray*}
The latter equality is just the normalization condition
(\ref{norm0}) of the wave function. Here the neutron momentum
$p^*_n$ is defined in the c.m. frame of the total reaction,
whereas the momentum $k$ is defined in the c.m. frame of the
$np$ system. So, $p^*_n$ and $k$ are the momentum of the same
particle, but in different frames. We replaced
$mp^{*2}_ndp^*_n d\Omega^*/\epsilon_{p^*_n}$ by
$\frac{md^3k}{\epsilon_k}$, since this integration volume is a
relativistic invariant.

The total cross section is obtained by integrating (\ref{sig2})
in the finite limits $\mee_{min}\le \mee\le \mee_{max}$, where
$\mee_{min}=2m_{e}$, $\mee_{max}=\sqrt{s_{\bar{p}d}}-m$.
Neglecting the electron mass, we can put $\mee_{min}=0$. For
$p_{\bar{p},lab}=1.5$ GeV/c the value $\mee_{max}$ is high
enough and provides the normalization condition (\ref{norm})
with very high accuracy.

To emphasize more distinctly the effect of the nuclear target,
we can represent the cross section (\ref{sig4}) of the
annihilation $ \bar{p}p \to e^+e^-$ on a free proton similarly
to eq. (\ref{sig2}):
\begin{equation}\label{sig5}
\frac{d\sigma_{\bar{p}p \to e^+e^-}}{d\mee}=\sigma_{\bar{p}p \to
e^+e^-}\delta(\mee-\sqrt{s_{p\bar{p}}})
\end{equation}
where $\sigma_{\bar{p}p \to e^+e^-}$ is defined in (\ref{sig4}),
$s_{p\bar{p}}=(p_p+p_{\bar{p}})^2= 2m^2+
2m\sqrt{p^2_{\bar{p},lab}+m^2}$.

The fact that in the annihilation on a free proton the mass of
the final $e^+e^-$ pair is fixed is reflected in (\ref{sig5}) in
the presence of the delta-function. Comparing this formula with
(\ref{sig2}), we see that the effect of the nuclear target
results in a dilation of the infinitely sharp distribution
$\delta(\mee-\sqrt{s_{p\bar{p}}})$ in a distribution of finite
width $\eta(\mee)$. The dilation of a distribution does not
change its normalization: $\eta(\mee)$ remains normalized to 1.

For a test of calculation we can find from (\ref{Mbar2}) (or
extract from (\ref{sig4})) the amplitude squared
$\overline{|M_{\bar{p}p\to e^+e^-}|^2}$, averaged over the angle
$\theta_{e^+}$:
$$
\overline{|M_{\bar{p}p\to e^+e^-}|^2}=
\frac{64\pi^2\alpha^2(\mee^2+2m^2)}{3\mee^2}
$$
Due to time invariance, it is equal to
$\overline{|M_{e^+e^-\to\bar{p}p}|^2}$. Then we can identify $m$
with the muon mass and find the total cross section of the
reaction $e^+e^-\to \mu^+\mu^-$:
$$
\sigma_{e^+e^-\to\mu^+\mu^-}=\frac{p^*\;
\overline{|M_{e^+e^-\to\mu^+\mu^-}|^2}}{64\pi {E^*}^3},
$$
($p^*$ and $E^*$ are the final muon momentum and energy). It
coincides with a well-known result and, at high $e^+e^-$ energy,
turns into
$\sigma_{e^+e^-\to\mu^+\mu^-}=\frac{4\pi\alpha^2}{3s_{e^+e^-}}$
given in textbooks. This confirms the
correctness of numerous factors appearing in our formulas.


\subsection{Analysis and numerical calculations}
\label{sec:theo-ana-num}

So far we have not defined the argument $k$ in the wave function
$\psi(k)$ determining the distribution (\ref{sig3}). The
integral over $dz$ incorporates an interval of $k$, which, as we
will see, for  $p_{\bar{p}}= 1500$ MeV/c and $\mee=2m$ starts
with the minimal value $k_{\min}\approx 360$ MeV/c. For these
values the relativistic effects become non-negligible, therefore
different relativistic approaches result, in principle, in
different expressions for $k$. If we take as the argument the
spectator momentum (the momentum of the neutron) in the rest
frame of the deuteron, then it is expressed as:
$k^2=\epsilon_n^2-m^2$, where the neutron lab. energy can be
represented as $\epsilon_n=$\mbox{$(p_d\cdot p_n)/M_d$}, so that
when $p_d=(M_d,\vec{0})$ we obtain identity. Introducing the
invariant $t=(p_d-p_n)^2$, and since \mbox{$p_d\cdot
p_n$}$=(M_d^2+m^2-t)/2$, we get:
\begin{equation}\label{q2}
k^2=\frac{(M_d^2+m^2-t)^2}{4M_d^2}-m^2.
\end{equation}
In the c.m. frame of the total reaction $\bar{p}d\to e^+e^-n$,
the variable $t$ is expressed as:
\begin{equation}\label{t}
t=(p_d-p_n)^2=M_d^2-2(\epsilon^*_d \epsilon^*_n-zp^*_d
p^*_n)+m^2
\end{equation}
where $\epsilon^*_d, \epsilon^*_n$ and $p^*_d, p^*_n$ are the
c.m. energies and momenta of the deuteron and the neutron,
$z=\cos\theta$ and $\theta$, as already mentioned, is the angle
between $\vec{p}^*_d$ and $ \vec{p}^*_n$.
The energy $\epsilon^*_n$ is given by eq. (\ref{En}), whereas
$\epsilon^*_d$ is obtained from (\ref{En}) by the replacement of
$\mee^2$ by $M^2_d$. After these precisions, the formula
(\ref{q2}) together with (\ref{t}) completely determines the
argument of the wave function. 

It should also be emphasized that the proton is
off-mass-shell. Its off-shell mass squared has the value
$(m^*)^2=p_p^2=t$, which depends on kinematics.

The minimal value $k_{\min}$ is achieved at
$z=1$, i.e. when the neutron recoils antiparallel to the
incident antiproton. Using eqs.(\ref{q2}) and (\ref{t}), 
one finds a value that is remarkably small:
 $k_{\min}\approx 360$ MeV/c, corresponding to an  off-shell mass
 $m^*\approx 800$ MeV ($\approx 0.85 m$) for the initial proton.

We emphasize the meaning of the integration over $z$ in eq.
(\ref{sig3}). The incident antiproton with momentum
$\vec{p}^*_{\bar{p}}$ in the $\bar{p}d$ c.m. frame meets in the
deuteron the proton with momentum $\vec{p}^*_p$. In the impulse
approximation (diagram of fig. \ref{eAMX}), the latter is
determined by the difference of the deuteron and the neutron
momenta: $\vec{p}^*_p=\vec{p}^*_d-\vec{p}^*_n$. Since eq.
(\ref{sig3}) gives the value of $\eta(\mee)$ at fixed $\mee$,
for all these momenta and the relative angles between them and
for the virtual proton mass squared $(m^*)^2=t$, the invariant
energy $\sqrt{s_{p\bar{p}}}=\sqrt{(p_p+p_{\bar{p}})^2}$ is just
equal to the given invariant mass $\mee$ of the virtual
$\gamma^*$ and of the final $e^+e^-$ pair. In this way, we
obtain the contribution of the events with given time-like value
of $q^2=\mee^2$ in the cross section. When the angle $\theta$
between $\vec{p}^*_n$ and $\vec{p}^*_d$ (and  $z=\cos\theta$)
varies, the modulus of the  proton momentum, its direction and
the off-shell mass $m^*$ vary correspondingly to provide the
same fixed value of $s_{p\bar{p}}=\mee^2=q^2$. The integral over
$z$ in (\ref{sig3}) incorporates all these events, with
different $\vec{p}^*_p,m^*$ and with the same $q^2$.

At first glance, the small near-threshold $\bar{p}p$ c.m. energy
$\mee\approx 2m$ in the collision of a fast $\bar{p}$
($p_{\bar{p}}=1500$ MeV/c) is achieved, when the antiproton
meets in the deuteron a fast proton having the same momentum as
the $\bar{p}$, in modulus and direction. The protons with such a
high momentum are very seldom in deuteron. For this mechanism,
the cross section would be very small. However, the
near-threshold value of $\mee$ is obtained in other kinematics.
As  we mentioned, the proton momenta $k$ in the deuteron wave
function $\psi(k)$ in eq. (\ref{sig3}) start with $k\approx
k_{min}\approx 360$ MeV/c only (that corresponds to  $z\approx
1$). The main reason which allows to obtain in this collision
the value $\mee\approx 2m$ is the off-shellness of the proton:
$m^*\le 0.85 m$ instead of $m^*=m$. This 15\%  decrease
relative to the free proton mass is enough to
obtain the invariant $p\bar{p}$ mass $\mee\approx 2m$, when one
has the two parallel momenta: 1500 MeV/c for $\bar{p}$ and 360
MeV/c for $p$.

We also notice that for other values of $z$, i.e. non-colinear
$p$ and $\bar p$ momenta, the proton momentum $k$ needed to
produce $\mee\approx 2m$ is larger than $k_{min}$. In a
correlated way, this proton is also further away from the
mass-shell.

The calculation in the framework of another relativistic
approach -- light-front dynamics
\cite{Carbonell:1998rj}
-- gives another formula for $k$ which numerically is
very close to the one of eq.(\ref{q2}),
reducing the minimal value $k_{\min}$ by 10 MeV/c only.

\begin{figure}[h!]
\begin{center}
\includegraphics[width=8cm]{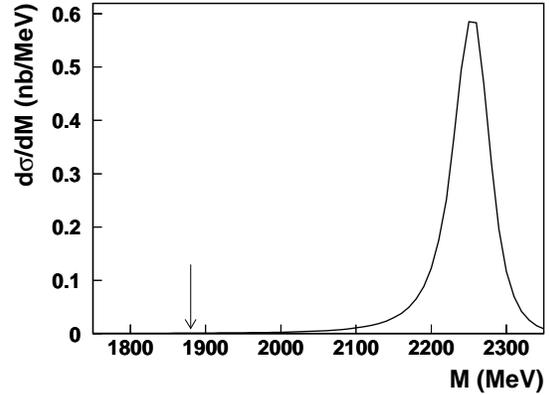}
\end{center}
\caption{The cross section $\frac{d\sigma_{\bar{p}d \to
e^+e^-n}}{d\mee}$ of the reaction $\bar{p}d\to e^+e^- n$ v.s.
$\mee$, in the interval: $1750\;MeV \leq \mee \leq 2350 \;MeV$,
calculated for a pointlike proton. The arrow indicates the $p
\bar p$ threshold.} \label{newplot2}
\end{figure}

The cross section $d\sigma_{\bar{p}d \to e^+e^-n}/d\mee$, eq.
(\ref{sig2}), has been calculated  for an antiproton of momentum
$p_{\bar{p}}=1500$ MeV/c on a deuteron nucleus at rest, with the
deuteron wave function \cite{Carbonell:1995yi}, incorporating
two components corresponding to S- and D-waves. The result is
shown in fig. \ref{newplot2}.
The maximum of the cross
section is at $\mee=2257$ MeV, that corresponds to the $\bar{p}$
interacting with a proton at rest (and on-shell).
The cross section integrated over $\mee$ is equal to 43 nb. We
remind that these calculations do not take into account the
proton form factor. Its influence will be estimated below. The
numerical integral over $\mee$ of the function $\eta(\mee)$, eq.
(\ref{sig3}), is $\approx 1$, in accordance with the
normalization condition (\ref{norm}).

The $p \bar p$ threshold value $\mee= 1880$ MeV is on the tail
of the distribution, far from the maximum. Relative to the
maximum, the cross section at threshold decreases approximately
by a factor 600. The numerical value at the threshold is:
\begin{eqnarray}
\left.\frac{d\sigma}{d\mee}\right|_{\mee=2m}=1\; \frac{pb}{MeV} \ .
\label{hf1pb}
\end{eqnarray}

\begin{figure}[h!]
\begin{center}
\includegraphics[width=8cm]{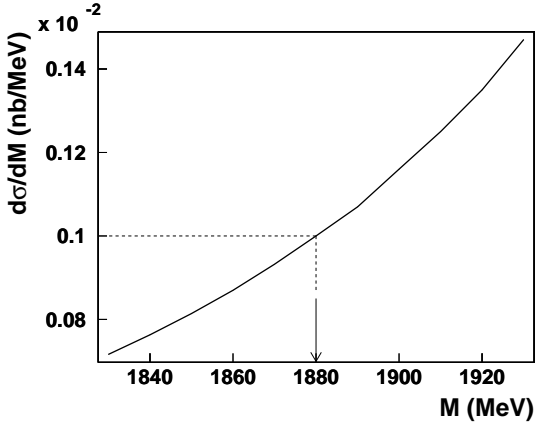}
\end{center}
\caption{The same as in fig.~\ref{newplot2}, but near the
$p\bar{p}$ threshold, in the interval $1830\;MeV \leq \mee \leq
1930 \;MeV$.}
 \label{newplot3}
\end{figure}

In fig.~\ref{newplot3} this cross section is shown in the
near-threshold interval $1830\;MeV \leq \mee \leq 1930 \;MeV$.
The integral over $\mee$ in a bin of width 100 MeV centered on
the threshold is:
$$
\int_{1830\;MeV}^{1930\;MeV}\frac{d\sigma_{\bar{p}d\to e^+e^-
n}}{d\mee}\;{d\mee}\approx 100\;pb \ .
$$

These estimations take into account the suppression
resulting from the momentum distribution in
deuteron. However, they do not incorporate the form factors
of the nucleon. To incorporate them in a simplified way, 
one can consider an effective form factor $\vert F \vert$ which
depends on $\mee$, and include it in the integral:
\begin{equation}
\sigma_{\bar{p}d\to e^+e^-n} = \int  \sigma_{\bar{p}p \to
e^+e^-}(\mee)\;\eta(\mee) \; \vert F(\mee) \vert^2 d \mee
\label{cswithff}
\end{equation}
where $\sigma_{\bar{p}p \to e^+e^-}(\mee)$ is the cross section for
pointlike nucleons given in eq.(\ref{sig4}).
To have an estimate of this  integral, we have taken the
effective proton form factor  measured in ref.~\cite{Aubert:2005cb}.
By doing this, we neglect all
off-shell effects.  We interpolate $\vert F(\mee) \vert$ linearly
between the measured values, and we limit the integral to the region
$\mee \ge 2m$. In this way we obtain
$\sigma_{\bar{p}d\to e^+e^-n} \simeq$ 1 nb, which is comparable to
the total cross section  $\sigma_{\bar{p}p
\to e^+e^-}$ on a free proton at $\mee=2257$ MeV.
\footnote{This cross section $\sigma_{\bar{p}p \to e^+e^-}=$ 1
nb at $\mee=2257$ MeV  is obtained from the measured cross
section $\sigma_{e^+e^- \to \bar{p}p}$  at the same c.m. energy,
which we estimate to be $\simeq 318$ pb,
from interpolation in the data of ref.~\cite{Aubert:2005cb}.
This value is then multiplied by the factor $\mee^2 / ( \mee^2 -
4m^2) \approx 3.27$ to obtain the cross section
in the inverse channel $\bar{p}p \to e^+e^- $ at the same c.m.
energy. }

\begin{figure}[h!]
\begin{center}
\includegraphics[width=8cm]{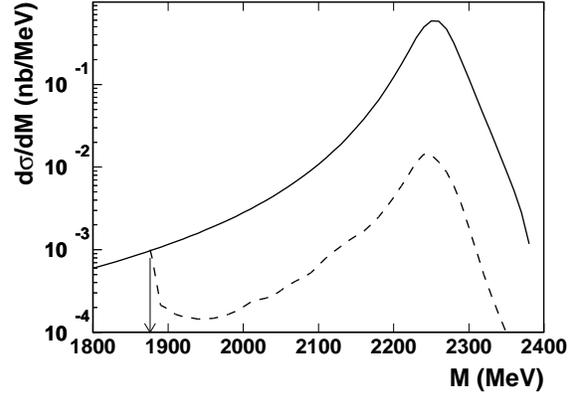}
\end{center}
\caption{ The cross section $\frac{d\sigma_{\bar{p}d \to
e^+e^-n}}{d\mee}$ including an effective form factor (see text)
is drawn as a dashed curve above the $p \bar p$ threshold. The
solid curve is the same as in fig.~\ref{newplot2} (but now in
logarithmic scale).}
 \label{newplot4}
\end{figure}

Figure~\ref{newplot4} shows the integrand of
eq.(~\ref{cswithff}) as a function of $\mee$, 
using the above choice of form factor. One
notices the rapid rise of the cross section as $\mee$ approaches
the $p \bar p$ threshold.
Lastly, we point out that at threshold, our differential cross
section $ d\sigma_{\bar{p}d \to n e+e-}$ of 1 pb/MeV
(eq.(\ref{hf1pb})) is not suppressed by any factor, since there
the form factor $\vert F \vert$ seems to be close to 1
experimentally
\cite{Baldini:2007qg,Baldini:2008nk}.  Below this threshold one
may expect a form factor effect {\it larger} than one, but this
is not known and this is the goal of the proposed study.

\subsection{Annihilation on heavier nuclei}
\label{sec:heavier-nuclei}

For $A>2$, we should  take into account the possibility of
excitation and breakup of the final nucleus $A-1$
in the process  $\bar{p}A\to (A-1)\gamma^*$. The result
contains the sum over the final energies of the residual nucleus
and the integral over a continuous spectrum. That is, the
function $|\psi(k)|^2$ in eq. (\ref{sig3}) is replaced by the
integral $\int_{E_{min}}^{E_{max}}S(E,k)dE$, where $S(E,k)$ is
the nucleus spectral function giving the probability to find in
the final state the nucleon with the relative momentum $k$ and
the residual nucleus with energy $E$. For high incident energy
we can replace the upper limit by infinity. Then we obtain:
$$
\int_{E_{min}}^{\infty}S(E,k)dE=n(k),
$$
where $n(k)$ is the momentum distribution in the nucleus.

To estimate the cross section  on heavy nuclei, we will still
use eqs. (\ref{sig2}), (\ref{sig3}) but with the two following
changes. ({\it i}) We replace the deuteron momentum distribution
by the nuclear one. Since the deuteron wave function is
normalized by (\ref{norm0}), whereas $n(k)$ is usually
normalized as $\int n(k)d^3k=1$, we replace $\psi^2(k)$ in
(\ref{sig3}) by $(2\pi)^3\epsilon_k n(k)/m$. ({\it ii}) We
multiply (\ref{sig3}) by the number of protons $Z$.

The numerical calculations were carried out for the 
$^{12}$C, $^{56}$Fe and $^{197}$Au  
nuclei with the nuclear momentum
distributions found in the papers
\cite{Antonov:2004xm,Antonov:2006md}.
Near threshold, i.e. at $\mee=1880$ MeV, for all three nuclei we
obtain very close results given by:
\begin{equation}\label{nucl}
\frac{d\sigma_{\bar{p}A\to e^+e^- X}}{d\mee}\;
\approx
6.5\,Z\,\frac{pb}{MeV}
\end{equation}
Multiplying by the charge $Z$ 
($Z(^{12}$C$  )=6 $, 
 $Z(^{56}$Fe$ )=26$, 
 $Z(^{197}$Au$)=79$) 
and integrating (\ref{nucl}) over
a 1 MeV interval near $\mee=1880$ MeV, we get:

$$
\sigma(^{12}\mbox{C})=39 \;pb,\;\; \sigma(^{56}\mbox{Fe})=0.17\;nb,
\;\; \sigma(^{197}\mbox{Au})=0.5\;nb.
$$

These results were found without taking into account the
absorption of $\bar{p}$ in nucleus before electromagnetic
annihilation. This absorption was estimated in Glauber approach
(it is applicable since now we do not need high nucleon
momenta). It reduces these cross sections by only a factor 2. We
recall again that these results are obtained for structureless
nucleons.


\subsection{Beyond the impulse approximation}
\label{sec:beyond}
 Without carrying out any calculation, we
discuss in this section other possible mechanisms for the
process $\pbard$. One of them is the initial state interaction,
which includes rescattering (not only elastic) of the initial
$\bar{p}$ in the target nucleus. In the rescattering, the
incident $\bar{p}$ looses energy and therefore the proton
momentum needed to form the invariant mass $\mee\approx 2m$
becomes smaller. The probability to find such a proton in
deuteron is higher. Therefore initial state interaction
increases the cross section.

Though, if the rescattering is inelastic (with pion creation),
it results in the reaction ${\bar p} \, d \to e^+ e^- N\pi$. In
this reaction, the $e^+ e^-$ pair is still produced in the
${\bar p}p$ electromagnetic annihilation, giving an information
about the proton timelike form factors. Therefore this process
is also interesting in itself, although it is not in the scope
of the paper.

Another mechanism of the reaction $\pbard$ may include also
the transition $\bar{p}p\to \bar{n}n$ with the subsequent
annihilation $\bar{n}n\to e^+e^-$. This transition can take
place, for example, via meson exchange:
\begin{eqnarray*}
&&\bar{p}p\to  (\bar{n}\pi^-)p  \to \bar{n}(\pi^- p)  \to
\bar{n}n,
\\
&&\bar{p}p\to \bar{p}(\pi^+ n) \to (\bar{p}\pi^+) n \to
\bar{n}n.
\end{eqnarray*}
If the annihilation $\bar{n}n\to e^+e^-$ occurs on the same
neutron which was created in the reaction $\bar{p}p\to
\bar{n}n$, this does not give anything new, since this is simply
a particular contribution $\bar{p}p\to \bar{n}n\to e^+e^-$ in
the initial state interaction incorporated in the full amplitude
$\bar{p}p\to e^+e^-$, i.e., in the timelike proton form factor.

On the contrary, when the annihilation $\bar{n}n\to \gamma^*$
$\to e^+e^-$ occurs on another neutron (the neutron from
deu\-te\-ron), then the reaction $\pbard$ gives information
about the timelike neutron form factor. So, what is measured is
the sum of two timelike form factors: the proton and the neutron
ones. One can expect that the direct annihilation $\bar{p}p\to
e^+e^-$ dominates over the mechanism with preliminary transition
$\bar{p}p\to \bar{n}n$. In this case, the contribution of the
proton form factor dominates.

We emphasize that in any case, whatever the intermediate steps
are in process (\ref{(1)}), the $e^+e^-$ pair of the final state
must come necessarily from the baryon-antibaryon electromagnetic
annihilation, $\bar{p}p$ or $\bar{n}n$, because there is only
one neutron left at the end. It cannot come from another
process, even if there are complicated intermediate steps, like
rescattering, etc. Therefore this $e^+e^-$ pair is a direct and
very little distorted probe of the baryon-antibaryon
electromagnetic annihilation vertex.

\section{Experimental aspects}
\label{sec-globalexp}

The aim of this section is to investigate the feasibility of
process (\ref{(1)}) in future experiments using antiproton
beams. The study is made in the case of the $\panda$ experiment,
which will use a state-of-the-art internal target detector at
the High Energy Storage Ring (HESR) at FAIR-GSI. This study is
only a first step, exploratory and rather qualitative. It
remains to be pursued in more details in the future.

Regarding the target nucleus, we restrict ourselves to the
deuteron. It is the only nucleus yielding a simple, three-body
final state $(e^+ e^- n )$,  for which an experimental strategy
can be defined. For heavier target nuclei $A$, the break-up
channels will dominate over the  three-body final state  \ $ e^+
e^- \ (A$-$1)$, and no simple identification strategy can be
defined. Indeed, to be able to use kinematical constraints in
order to select the desired reaction, one needs to detect (at
least) all final particles but one, a task which becomes more
and more difficult as the number of nuclear fragments increases. 
As a side remark, we also note that the HESR luminosity in
$({\bar p} A)$ decreases with the atomic charge
Z of the target nucleus~\cite{Panda:2009}, in a way that 
roughly compensates the increase of  cross section with Z 
reported in sect.~\ref{sec:heavier-nuclei}.

New technologies for antiproton accelerators, targets and
detectors provide a gain in luminosity and allow to access
processes with very low cross sections, like the one presented
here in the near-threshold region. In this section we first 
examine reaction (\ref{(1)})  (the {\it signal}) and then the 
competing background processes.
The presented material largely relies on the design performances 
of the $\panda$ detector 
\cite{Panda:2009,Panda:2005} and the study of the electromagnetic process 
 $\pbarp$ on a free proton at rest~\cite{Panda:2009,Panda:2005,Sudol:2009vc}.

\subsection{Beam and target operating conditions}
\label{sec:opcond}

Considering the process ${\bar p} + p_{bound} \to \gamma^*$ on
an off-shell proton bound in the nucleus, the lower the
antiproton beam momentum, the easier it is to reach 
a photon virtuality $Q^2 < 4m^2$ $(Q^2=q^2)$. 
Ideal conditions would be antiprotons of  very low
momentum. Here we consider the nominal lowest beam momentum that
is foreseen in the HESR, i.e.  $p_{{\bar p},lab}=$ 1.5 GeV/c. As seen in 
sect.~\ref{sec:theo-ana-num}, in these conditions only a very
small fraction of the target protons in the deuteron will have
enough off-shellness in order to produce a ${\bar p} p$ system
of invariant mass squared $\le 4m^2$. In the impulse
approximation, i.e. assuming that the neutron is a spectator in
the  $\pbard$  process,  these off-shell protons are found in the
upper tail of the momentum distribution $n(k)$ of the deuteron,
at $k=360$ MeV/c and above.

In the $\panda$ experiment, nuclear targets will be used to study
hadrons properties in the nuclear medium. For deuterium, an
internal target of cluster-jet type or pellet type will be
placed in the beam, and an effective target thickness of $3.6
\cdot 10^{15}$ atoms/cm$^2$ is considered. At $p_{{\bar
p},lab}=1.5$ GeV/c, beam losses limit the  number of antiprotons
to  $10^{11}$ per cycle, yielding a maximal value of  $5 \cdot
10^{31}$ cm$^{-2}$.s$^{-1}$  for the cycle average luminosity
${\cal L}$ in $( {\bar p} d)$ collisions \cite{Panda:2009}.

\subsection{Count rate estimate for the process  $\pbard$ }
\label{sec:rate}

With the above operating conditions ($p_{{\bar p},lab}$, 
${\cal L}$) and the cross section $\sigma$ calculated in 
sect.~\ref{sec:theo-ana-num}, one can compute the number of
events from process (\ref{(1)})  in an  ideal detector, covering
the  $4 \pi$ solid angle with 100 \% efficiency;  it is $N=
{\cal L} \cdot \sigma$ . For one month effective beamtime
(i.e. corresponding to an integrated luminosity equal to 
${\cal L} \times 2.6 \cdot 10^6 s$),
this yields a large number of events integrated over the full
range of dilepton invariant mass: $N\sim 1.3 \times 10^5$ (this
number is deduced from the cross section $\sigma_{\bar{p}p \to
e^+e^-} \sim $ 1 nb on a free proton at rest). However, in the
special region of interest near the $N {\bar N}$ threshold, the
cross section $d\sigma / d \mee $ is much reduced, of the order
of 1pb/MeV (see sect.~\ref{sec:theo-ana-num}), yielding a rate
of 130 events per month in a 1 MeV bin of dilepton mass. This
number of events is small, but other rare processes considered
in the $\panda$ physics program have comparable expected rates.

On the one hand, the global detector efficiency  and the
necessary experimental cuts will reduce this amount of good
events, with a reduction factor that could reach 50\%. On the
other hand, the cross section calculated in 
sect.~\ref{sec:theo-ana-num} for process (\ref{(1)}) is
probably a lower limit in the subthreshold region, since there
the proton form factor can be larger than one and enhance the
cross section. Therefore, all in all, within one month effective
beamtime one could obtain an experimental spectrum of dilepton
mass $\mee$ containing $\sim$ a hundred signal events per MeV
bin near the threshold ($\mee \sim 2m$). This would allow  a
first insight into possible structures in this region, including
the totally unexplored subthreshold region.

\subsection{Experimental signature of the exclusive channel
$\pbard$ and related background processes}
\label{sec:signature}

The selection of the exclusive reaction (\ref{(1)}) is based on
the detection of the lepton pair. The experimental strategy
relies on three main characteristics of the detector: 1) its
resolution, 2) its hermeticity and 3) its particle
identification capability. \vskip 2 mm

Regarding the first point, the main variable is the missing mass
$M_X$, i.e. the invariant mass of the missing system $X$ in the
reaction ${\bar p} \, d \to e^+ e^- X$. It is defined using
the four-momentum vectors of the initial and detected particles:
\begin{eqnarray*}
M_X^2 \ = \ ( \  p_{\bar p} + p_d - p_{e^+} - p_{e^-} \ ) ^2 \ .
\end{eqnarray*}
Due to the many processes producing a lepton pair inclusively,
this missing mass will have a wide distribution; for the events
of reaction (\ref{(1)}) a peak must be searched at the neutron
mass, which represents the physical lower bound of the $M_X$
spectrum. Therefore the experimental resolution in this missing
mass is a crucial parameter (see sect.~\ref{sec:simul}).

The background due to inclusive lepton pair production is quite
large  at the raw level. True $e^+e^-$  pairs originate mostly
from  Dalitz decay (or direct $l^+l^-$ decay) of mesons, and
also from real photon conversion in the target. These types of
processes will give dileptons mostly at low invariant mass 
($\mee \le 1$
GeV), while we are looking for high-mass ones ($\mee \ge 1.8$
GeV). Uncorrelated pairs coming from combinatorial background
may form a high invariant mass; however they can be subtracted
using the like-sign pairs (also combinatorial) as in heavy-ion
dilepton experiments.
All these processes of inclusive lepton pair production
correspond  to missing masses $M_X$ larger than one neutron
mass, and a large fraction of them can be eliminated by a proper
cut in $M_X$. Also, these processes create more than three
particles in the final state, and a condition on the observed
particle multiplicity should be efficient to reject a large
fraction of this background.

\vskip 2 mm

This brings us to the second important aspect, of detector
hermeticity. The $\panda$ detector will cover almost the 4$\pi$
solid angle, by combining a target spectrometer for large polar
angles and a forward spectrometer for small angles. The good
hermeticity is essential in the physics program in order
to detect a complete spectrum of final states. In reaction
(\ref{(1)}) the neutron will go undetected most of the times;
actually, in our kinematics of interest near threshold, this
neutron is emitted at very backward angle w.r.t. the beam and
will not hit any detector. With respect to hermeticity,
potential background to reaction (\ref{(1)}) consists in
reactions in which, in addition to the $e^+e^-$ pair, one or
more light particles (pions, photons) are emitted and escape
detection. They will contribute to form a missing system $X$ (as
defined above) of low baryonic mass, typically $M_X \simeq
m+m_{\pi}$, polluting the region of interest. The simplest cases
of such background reactions are:

\centerline{${\bar p} \, d \to e^+ e^- n \, \pi^0$ \ \ \ \
(from \ ${\bar p} \, p_{bound} \to \gamma^*  \, \pi^0$)       }
\centerline{${\bar p} \, d \to e^+ e^- p \, \pi^-$ \ \ \ \
(from \ ${\bar p} \, n_{bound} \to \gamma^*  \, \pi^-$)  }
\noindent on the bound proton or the bound neutron in the
deuteron. The  final $\pi N$ state can be non-resonant or
resonant (e.g. the $\Delta(1232)$).

\vskip 2 mm A third important aspect is the particle
identification capability of the detector (PID), for charged and
neutral particles. In $\panda$, a global
estimator~\cite{Panda:2009} combines the information of the
various subdetectors and will provide charged particle
identification in the momentum range from 200 MeV/c to 10
GeV/c. In our case the most important task is to discriminate
between $e^{\pm}$ and $\pi^{\pm}$. Indeed, the main competing
background to reaction (\ref{(1)}) is of the type 
${\bar p} \, d \to \pi^+ \pi^- n $ 
or ${\bar p} \, d \to \pi^+ \pi^- X $
where the $X$ system has a low baryonic mass, and the $\pi^+
\pi^-$ pair is misidentified as an $e^+ e^-$ pair.

The case of $\pbarp$ on a free proton at rest has been studied
in detail~\cite{Sudol:2009vc}  in view of future measurements of 
the timelike proton form factors in $\panda$.
The exclusive process
${\bar p} p \to \pi^+ \pi^-$ is the dominant background to
this measurement, with a cross section $\sim 10^6$ larger than
the $\pbarp$ cross section in the HESR energy range.
Simulation studies 
have been performed in order to estimate precisely the fraction of 
${\bar p} p \to \pi^+ \pi^-$  events that can be eliminated by PID
cuts. The most efficient elements for $\pi^{\pm} / e^{\pm}$
discrimination are the electromagnetic calorimeters and the
$dE/dx$ measurements in the tracking systems. Applying the
global PID cut at the ``very tight'' level 
\footnote{
corresponding to a global identification probability greater 
than 99.8\% for an $e^+$ or an $e^-$.
}
yields an efficiency
to single electrons of $\sim$ 90\% with a contamination rate of pions
smaller than $10^{-3}$, for particle momenta above 1
GeV/c~\cite{Panda:2009}. As a result, a rejection factor of
$10^{-7}$ is reached on the pion pair of the process 
${\bar p} p \to \pi^+ \pi^-$. This background is further reduced by a
factor 50-100 when performing a two-body kinematical fit. In this way 
one obtains a more than 99\% clean sample of $\pbarp$ events, 
with a 20-30\% global efficiency~\cite{Sudol:2009vc}.

These results cannot be extrapolated directly to the case of
reaction (\ref{(1)}), for which a detailed simulation would be
required. Nonetheless, the strategy is based on the same
elements. In a simplified scheme, the use of the ``very tight''
PID cut allows to reduce the 
${\bar p} \, d \to \pi^+ \pi^- n$  
hadro\-nic background by a factor $\sim 10^6$ ($10^3$ for
each charged pion). After that, one is left with a
back\-ground-to-signal ratio of about 1:1. Kinematical
constraints are less efficient than in the case of $\pbarp$
since we have a three-body final state. Here only one constraint
can be imposed, on the missing mass. Therefore the remaining
background from hadronic processes cannot be eliminated by a
cut, it has to be calculated and subtracted. It is in principle
possible, provided that the cross section for these processes is
measured (in $\panda$) and the PID performan\-ces of the detector
are known with accuracy. Other reactions due to strong ${\bar p}
n$ annihilation are also potential sources of background. For
example the process ${\bar p} n \to \pi^- \pi^0$ followed by
$\pi^0$ Dalitz decay, creates a $\pi^- e^+$ pair that can be 
misidentified as an $e^- e^+$ pair; etc.


In summary, the simulations already performed for the channel
$\pbarp$  serve as a benchmark study to explore the possibility
to measure the process $\pbard$ in $\panda$. The competing
background to this reaction is certainly more important than in
the case of $\pbarp$  on a free proton at rest, and
 more difficult to reject. We have given elements for a strategy, but
the full experimental feasibility remains to be proven. A
further study of this point will require complex simulations
which are beyond the scope of this paper.

\subsection{First-order simulation of \  $\pbard$ }
\label{sec:simul}

\begin{figure}[htb]
\begin{center}
\includegraphics[width=8.3cm]{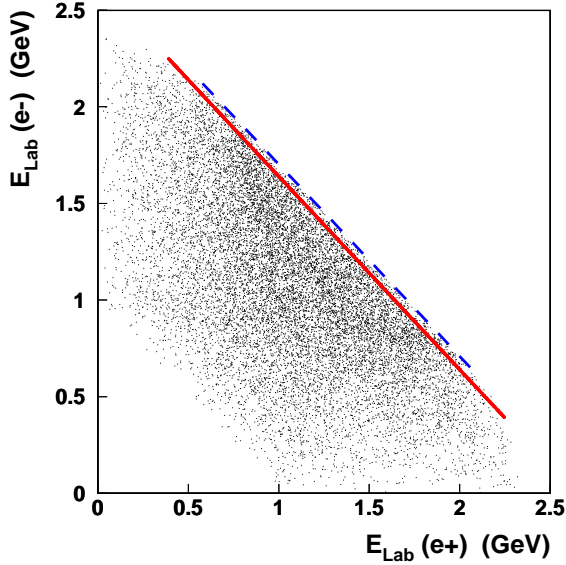}
\end{center}
\caption{ Simulation of the reaction  $\pbard$ at  antiproton
beam momentum $p_{{\bar p},lab}=$ 1.5 GeV/c: electron energy
versus positron energy in the Lab frame. The solid line
corresponds to our threshold kinematics of interest (see text).
The dashed line corresponds to the reaction $\pbarp$ on a free
proton at rest, at the same value of $p_{{\bar p},lab}$.}
\label{figexp01}
\end{figure}

A simple Monte-Carlo simulation of reaction (\ref{(1)}), of pure
phase-space  type, was performed. The detector resolution is
implemented in order to estimate some important parameters, such
as the achievable resolution in missing mass ($\sigma_{M_X}$) 
and in dilepton mass ($\sigma_{\mee}$).

Kinematics are defined by an incoming antiproton of 1.5 GeV/c
momentum onto a deuteron nucleus at rest. The two-body reaction
\ ${\bar p} \, d \to n \gamma^*$ \ is generated by sampling in
$\cos \theta_{\gamma^* cm}$ (the polar angle) and in the
azimuthal angle $\phi_{\gamma^* cm}$ in the \ ${\bar p}  d$ \
center-of-mass. The photon virtuality $Q^2=( \mee )^2$ is
sampled between $Q^2_{min}=4 m_e^2$ and  $Q^2_{max}= ( \sqrt{
s_{{\bar p} p}} - m )^2$. Then the decay $\gamma^* \to e^+ e^-$
is generated in the $\gamma^*$ center-of-mass by sampling in
$\cos \theta_{e^-}$ and $\phi_{e^-}$. Finally all the produced
particles are transformed back to the laboratory frame. The five
samplings are made in uniform distributions, therefore the event
weight is not realistic for this first study.

\begin{figure}[htb]
\begin{center}
\includegraphics[width=8.3cm]{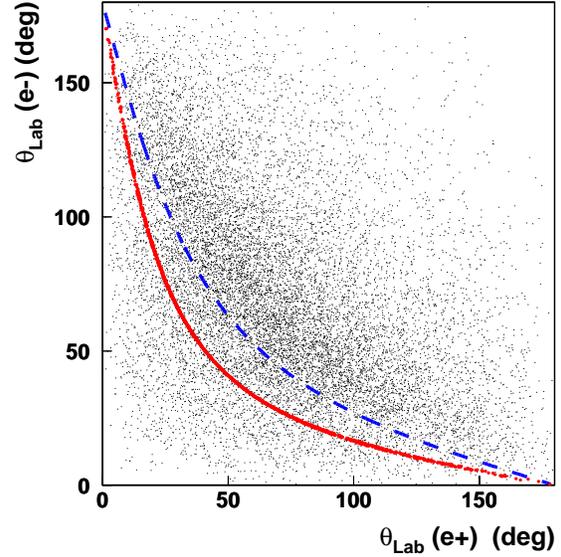}
\end{center}
\caption{Simulation of the reaction  $\pbard$ at $p_{{\bar
p},lab}=$  1.5 GeV/c: polar angle of the electron versus polar
angle of the positron in the Lab frame. The solid curve
corresponds to our threshold kinematics of interest (see text).
The dashed curve corresponds to the reaction $\pbarp$  on a free
proton at rest, at the same value of $p_{{\bar p},lab}$. }
\label{figexp02}
\end{figure}

The detector  resolution is implemented on the momentum and
angles of the electron and positron produced at the vertex, by
sampling a measurement error in a Gaussian distribution,
independently for $(p, \theta, \phi)_{Lab}$ of each particle.
For the relative momentum resolution ($\sigma_p /p$) we take a
uniform value of 1.5\% in r.m.s., which should be conservative 
given the rather low momenta involved ($p \le 2.2$ GeV/c). 
The angular resolution is taken to be     
$\sigma_\theta$ = $\sigma_\phi = 3 mr$ in r.m.s.

The resulting  phase space of the lepton pair is shown in
fig.~\ref{figexp01} for the energies and in 
fig.~\ref{figexp02}
for the polar angles. When compared to the case of the reaction
$\pbarp$  on a free proton at rest (dashed curves), now most of
the correlations between the two leptons are lost and the phase
space is much more open. The special threshold kinematics of
interest in reaction (\ref{(1)}) is defined by: $i$) a photon
virtuality corresponding to the $N \bar N$ threshold, i.e. $\mee
=2m$ , and $ii$) the virtual photon emitted at forward angle in
the lab (cf. sect.~\ref{sec:theo-ana-num}: this corresponds to
$z=1$ or $k=k_{min}$ in the deuteron momentum distribution
$n(k)$). For this particular kinematics, represented by a solid
curve on Figs.~\ref{figexp01} and~\ref{figexp02},
lepton-to-lepton correlations are re-established, but they
differ from the case of a free proton at rest, because the
target proton is now off-shell in the deuteron.

\begin{figure}[htb]
\begin{center}
\includegraphics[width=8.3cm]{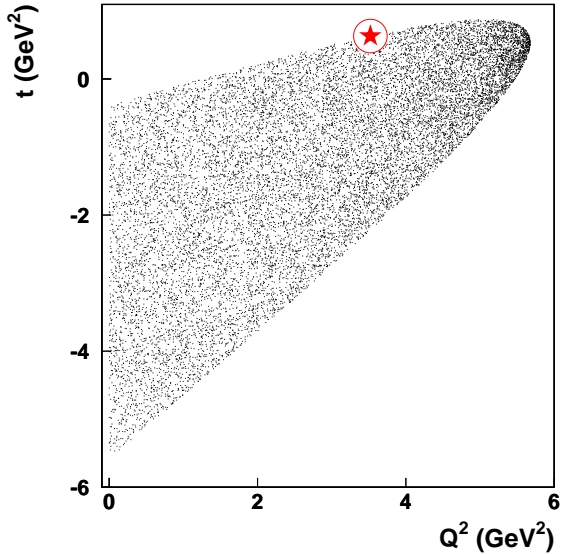}
\end{center}
\caption{Simulation  of the reaction  $\pbard$ at $p_{{\bar
p},lab}=$ 1.5 GeV/c: the variable $t$ (see text)
versus the photon virtuality $Q^2 = ( \mee )^2$. The circled
star corresponds to our threshold kinematics of interest, where
$t=0.63$ GeV$^2$.}  \label{figexp03}
\end{figure}

The off-shellness of the initial proton in process (\ref{(1)})
has been defined in sect.~\ref{sec:theo-ana-num}. In the
impulse approximation, the off-shell mass squared of this proton
is equal to  $t=(p_d - p_n )^2$, where $p_d$ and $p_n$ are the
four-momentum vectors of the initial deuteron and final neutron.
In this approximation, the proton off-shellness is an
experimentally accessible quantity since the neutron is
``measured'' as the missing particle. Figure~\ref{figexp03}
shows the $t$ variable as a function of the photon virtuality,
in the full phase space. In the threshold kinematics of
interest, $t$ is equal to 0.63 GeV$^2$ (to be compared with the
free proton case,  $t=M_p^2=$ 0.88 GeV$^2$).

\begin{figure}[htb]
\begin{center}
\includegraphics[width=8.3cm]{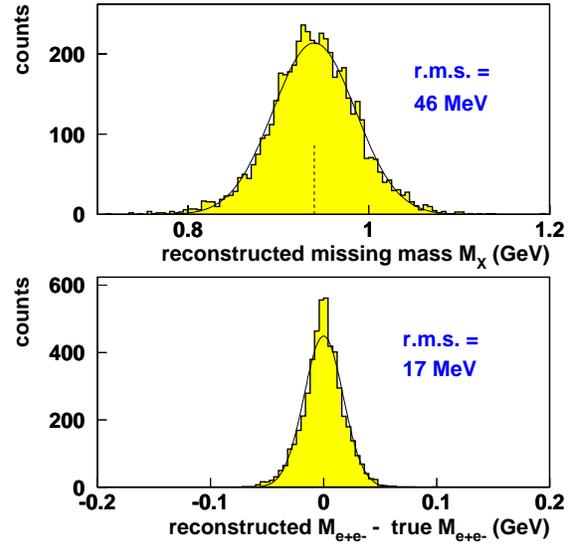}
\end{center}
\caption{Simulation of the  reaction  $\pbard$ at $p_{{\bar
p},lab}=$ 1.5 GeV/c: resolution obtained for the missing mass
$M_X$ (upper plot) and for the dilepton invariant mass $\mee$
($\equiv \mathrm{M}_{e^+e^-}$, lower plot).} \label{figexp04}
\end{figure}

Finally, fig.~\ref{figexp04} shows  the experimental
resolution that can be expected on the most important variables:
the missing mass $M_X$ and the dilepton invariant mass $\mee$.
These resolutions are quite uniform within the reaction phase
space, apart from a slight $Q^2$-dependence. The width of the
distributions in fig.~\ref{figexp04} is dominated by the
momentum resolution of the detected $e^+ e^-$ pair. The obtained
resolution in missing mass is rather good, with an   
r.m.s. of 46 MeV, or  a
full-width at half-maximum of 106 MeV. This value is 
smaller than one pion mass, therefore by applying a cut around the
neutron mass in the $M_X$ spectrum one should be able to
separate - at least partly - the missing systems $X$ of the type
(one nucleon + one pion), which are the closest in the spectrum.

The resolution in dilepton invariant mass (r.m.s.= 17 MeV) is also
reasonably good, and well suited to evidence some structures due
to possible baryonium bound states near the $p \bar p$
threshold.

\section{Other aspects}
\label{sec-other}

In this paper we have concentrated mostly on
the possibility to use process (\ref{(1)}) to access the $p \bar
p$ threshold and under-threshold region. However, we would like
to stress that this process is also interesting for other
purposes.

First, it provides a full range of $e^+e^-$ invariant mass
$\mee$ with one single antiproton beam energy. In this sense,
the role of the nucleus is a bit similar to the role of initial
state radiation in the inverse reaction 
$e^+e^- \to p \bar p \gamma_{ISR}$. 
As a consequence, any clean experimental data on process
(\ref{(1)}) at any beam energy (e.g. HESR momenta higher than
1.5 GeV/c), have the potential to yield valuable information on
the (half off-shell) nucleon timelike form factors, in a large
$Q^2$-range. Of course, in this perspective more theoretical
work is needed to test the validity of the impulse
approximation, and to better define the off-shell effects in the
$N {\bar N} \gamma^*$ vertex.

Second, one should note that the antiproton momentum 1.5 GeV/c
just corresponds to the threshold value of creation of the
$\Lambda\bar{\Lambda}$ pair on a free proton. Therefore the
virtual creation of the $\Lambda\bar{\Lambda}$ pair in reaction
(\ref{(1)}) is not suppressed by the nucleon momentum
distribution in deuteron and contributes just in the domain of
the peak of fig.~\ref{newplot2}, that allows one to study the
$\Lambda\bar{\Lambda}$ threshold region with good statistics. In
the $\Lambda\bar{\Lambda}$ system, the quasi-nuclear states were
predicted in \cite{Carbonell:1993dt} and, similarly to the
$N\bar{N}$ quasi-nuclear states, they should manifest themselves
as irregularities in the cross section. The contribution of the
channel $\bar{p}p\to \bar{\Lambda}\Lambda\to e^+e^-$ in the
total cross section $\bar{p}p\to e^+e^-$ (the latter equals 1
nb, see sect.~\ref{sec:theo-ana-num} above) was estimated 
in~\cite{Dalkarov:private} as 0.1 nb, 
i.e. 10\% of the total cross section. Therefore we expect that the
structures caused by the channel $\bar{p}p\to
\bar{\Lambda}\Lambda\to e^+e^-$ can be observed in process (1) 
in the region of mass $\mee$ near the $\Lambda\bar{\Lambda}$ threshold.

\section{Conclusion}
\label{sec-concl}

We have studied the reaction  $\bar{p}A\to (A-1)\gamma^*$
(followed by $\gamma^*\to e^+e^-)$. This process gives access to
the ${\bar p} p$ annihilation $\bar p p \to \gamma^*$ at
invariant masses $\sqrt s_{{\bar p} p}$ which are below the
physical threshold of  $ 2 m$, due to the proton off-shellness in
the nucleus. In this way a possibility exists to access the
proton timelike form factors in the near-threshold and the
totally unexplored under-threshold region, where  $N \bar N$
bound states are predicted.

The differential cross section $d\sigma / d \mee $ has been
calculated as a function of the dilepton invariant mass $\mee$,
for an incident antiproton of 1.5 GeV/c momentum on a deuteron
target (and heavier nuclei). The calculation is done in the
impulse approximation; then
the distribution of $\mee$  is obtained from the deuteron wave
function $\psi(k)$. We find that the
$p \bar p$ threshold ($\mee=2m$) is reached for a minimal proton
momentum $k_{\min}$=360 MeV/c in the nucleus, and at this point
the cross section is about 1 pb/MeV. The calculation does not
include the form factor effect, which should come as an extra
factor.

Experimental aspects have been investigated in the case of a
deuteron target, i.e. for the three-body process $\pbard$. We
have taken the conditions of the $\panda$ project at FAIR-GSI: an
antiproton beam momentum of 1.5 GeV/c and the detection of the
lepton pair. The count rate in the near-threshold region of
$\mee$ is small but not negligible. The main difficulty is to
identify the reaction among the hadronic background which is
about six orders of magnitude higher. First elements of strategy
were presented for this background rejection, based on particle
identification, detector hermeticity, and missing mass
resolution. Although the subject would require a much more
detailed study, we conclude that this process has a chance to be
measurable in $\panda$, given the very good design performances of
the detector.

\section*{Acknowledgements}
We would like to thank the GDR-Nucl\'eon (n$^o$ 3034) for
providing the infrastructure for this work and financial
support. This work is  supported in part by the
French CNRS/IN2P3. \\
The authors are indebted to O.D.~Dalkarov for his interest to
this work, for stimulating discussions and useful advices 
and to J.-F. Mathiot for his support.
\\
V.A.K. is sincerely grateful for the warm hospitality of the
Laboratoire de Physique Corpusculaire, Universit\'e Blaise
Palcal, Clermont-Ferrand, France, where part of this work was
performed. V.A.K. is indebted to A.N. Antonov for illuminating
discussions and for kind\-ly providing the data on the nuclear
momentum distributions \cite{Antonov:2004xm,Antonov:2006md}.
\\
H.F. would like to thank the $\panda$ Group at IPN-Orsay, 
and especially T.Hennino, for useful experimental advices.

\bibliography{ref222}

\end{document}